\newcommand{\C}{{\Bbb C}}
\newcommand{\ket}[1]{{|#1\rangle}}

\documentstyle[aps,multicol,amssymb,eqsecnum]{revtex}
\begin{document}
\draft

\title{Comment on `Decomposition of pure states of a quantum register'}
\author{A. Yu. Vlasov}
\address{FRC/IRH, 197101, Mira Street 8, St.-Petersburg, Russia}
\date{10 November 2000}
\maketitle
\begin{abstract}
 I. Raptis and R. Zapatrin in the quant-ph/0010104 show possibility to
express general state of $l$-qubits quantum register as sum at most
$2^l-l$ product states. In the comment is suggested more simple construction
with possibility of generalization for decomposition of tensor product of
Hilbert spaces with arbitrary dimension $n$ (here simplicial complexes
used in the article mentioned above would not be applied directly).
In this case it is decomposition with $n^l-(n^2-n)l/2$ product states.
\end{abstract}

\begin{multicols}{2}

\section{Decomposition of qubits and qu$n$its}

 Result of \cite{RZ} is proved with using simplicial complexes, etc.,
but it can be shown also with less special constructions. Such simplification
may be useful because it makes possible to apply similar
construction not only for qubit spaces $\C^2 \otimes \cdots \otimes \C^2$,
but also for `qu$n$it' spaces $\C^n \otimes \cdots \otimes \C^n$.
Here is represented only short introduction to construction with arbitrary
$n$, because {\em the result is already known and described in }\cite{Sudbery}.

The result of \cite{RZ} is related with possibility of exclusion
by local unitary transformations of $l$ terms with one `unit' like
$\ket{100\ldots0}$, $\ket{010\ldots0}$, etc., from general $l$-qubit state
with $2^l$ terms.

 To show it, let us consider some state
$\ket{h} = \alpha\ket{00\ldots0} + \alpha_1 \ket{10\ldots0} + \cdots$,
where $\alpha$ may be zero. It is possible with using a local unitary
transformation in first qubit
$U_1\colon (\alpha,\alpha_1) \to (\alpha',0)$ to produce state
$\ket{h'} = \alpha'\ket{00\ldots0} + 0 \ket{10\ldots0} + \cdots$,
where $|\alpha'|^2 = |\alpha|^2 + |\alpha_1|^2$. Next, because
$\ket{h'} = \alpha'\ket{00\ldots0} + \alpha_2 \ket{01\ldots0} + \cdots$,
it is possible to eliminate $\alpha_2$ by a local unitary transformation
of second qubit and with $l$ similar steps we can produce some state
$\ket{h^{(1)}}$. If $\ket{h}$ is product state, then $\ket{h^{(1)}}$
has only one nonzero term, but in general case it is possible to
produce sum with no more than $2^l-l$ terms only by iterating the process:
$\ket{\tilde{h}}=\lim_{N \to \infty} \ket{h^{(N)}}$.

To check convergence let us consider a slightly different algorithm, when
$N$'th step is elimination of term with one unit in position $i_N$ where
coefficient $\alpha_{i_N}^{(N)} = \alpha_{\max}^{(N)}$ is maximal
between $l$ similar terms. Absolute value
of coefficient $\alpha^{(N)}$ of term $\ket{00\ldots0}$ is always limited
$|\alpha^{(N)}| \le |h|\equiv 1$ and meets an equation:
$|\alpha^{(N)}|^2 = |\alpha|^2+\sum_{K=1}^N |\alpha_{\max}^{(K)}|^2$.
So $\sum_{N=1}^\infty |\alpha_{\max}^{(N)}|^2 = |\tilde\alpha|^2 - |\alpha|^2$
and then $\lim\limits_{N \to \infty} \alpha_{\max}^{(N)}~=~0$.

To generalize the method to arbitrary $n > 2$ (`qu$n$it' \cite{qunit})
let us consider first $n = 3$ (`qutrit' \cite{qutrit}). If we start with some
state $\ket{h} = \alpha\ket{00\ldots0} + \alpha_1 \ket{10\ldots0} +
\beta_1 \ket{20\ldots0} + \cdots$, it is possible with using a local unitary
$SU(3)$ transformation in first qutrit
$U_1\colon (\alpha,\alpha_1,\beta_1) \to (\alpha',0,0)$ to produce state
$\ket{h'} = \alpha'\ket{00\ldots0} + 0 \ket{10\ldots0} +
0 \ket{20\ldots0} + \cdots$ with $|\alpha'|^2 = |\alpha|^2 + |\alpha_1|^2
+ |\beta_1|^2$ and algorithms similar with discussed above may eliminate
$2l$ components with one `1' or `2' and $l-1$ `0'.

But in the case it is not all components appropriate for exclusion.
Let us write state $\ket{g}$ without the $2l$ term as:
$\ket{g} = \gamma\ket{11\ldots1} + \gamma_1 \ket{21\ldots1} + \cdots$.
Because it is possible to transform qutrit space without change of
component $\ket0$, i.e.,
$U_1\colon (\alpha,\gamma,\gamma_1) \to (\alpha,\gamma',0)$ we can
exclude all $l$ components with one `2' and $l-1$ `1' and so it is
possible to have no more than $3^l-3l$ components.

The similar process is possible to apply to arbitrary qu$n$it $\C^n$.
First, we can exclude $(n-1)l$ terms with one `1' and $l-1$ `0'.
Second, $(n-2)l$ terms with one `2' and $l-1$ `1'. After $n-1$
iterations we can neglect $((n-1)+(n-2)+\cdots+1)l = n (n-1) l /2$ terms
and so we have no more than
$$n^l - \case1/2 n (n-1) l$$
components. For $l=2$ we have $n^2 -  n (n-1) 2 /2 = n$ terms of
usual Schmidt decomposition.

\section*{Acknowledgements}
Author is grateful to R.~Zapatrin and C.~Zalka for explanations and comments.
Many thanks to A.~Sudbery for reference to the work \cite{Sudbery}.

\end{multicols}
\end{document}